\documentclass[superscriptaddress,prl,final,twocolumn,showpacs,showkeys]{revtex4}
\usepackage{amsmath}
\usepackage{graphicx}
\usepackage{amsfonts}
\usepackage{amssymb}
\usepackage{subfigure}
\begin{document}

\title{Tunneling phase gate for neutral atoms in a double-well lattice}
\author{Frederick W.\ Strauch}
\email{frederick.strauch@nist.gov}
\affiliation{Department of Physics, Gettysburg College, Gettysburg, PA, 17325, USA}
\affiliation{National Institute of Standards and Technology, 
Gaithersburg, MD 20899, USA
}
\author{Mark Edwards}
\affiliation{Department of Physics, Georgia Southern University, Statesboro, GA, 30460--8031, USA}
\affiliation{National Institute of Standards and Technology, 
Gaithersburg, MD 20899, USA
}
\author{Eite Tiesinga}
\author{Carl Williams}
\author{Charles W.\ Clark}
\affiliation{National Institute of Standards and Technology, 
Gaithersburg, MD 20899, USA
}
\date{\today}

\begin{abstract}
We propose a new two--qubit phase gate for ultra--cold 
atoms confined in an experimentally realized tilted double--well optical 
lattice [Sebby--Strabley, {\it et al.}, \pra {\bf 73} 033605 (2006)].  Such a 
lattice is capable of confining pairs of atoms in a two--dimensional array 
of double--well potentials where control can be exercised over the barrier 
height and the energy difference of the minima of the two wells (known as 
the ``tilt'').  The four lowest single--particle motional states consist of two pairs of motional states in which each pair is localized on one side of the central barrier, allowing for two atoms confined in such a lattice to be spatially separated qubits.  We present a time--dependent scheme to manipulate the tilt to induce tunneling oscillations which produce a collisional phase gate. Numerical simulations demonstrate that this gate can be performed with high fidelity.

\end{abstract} 
\maketitle

Quantum information processing with neutral atoms in optical lattices holds great promise, but achieving the necessary single-qubit and two-qubit control remains an elusive experimental challenge.  Recent experimental work has made great progress on this front by developing techniques to isolate and manipulate an ensemble of pairs of atoms in a double-well lattice \cite{Anderlini2006,Jenny2006}.   Previous theoretical proposals for quantum gates include direct collisional interactions in state-dependent potentials \cite{Jaksch99,Calarco2000}, long-range dipole-dipole interactions \cite{Brennen99}, and Rydberg states for maximal speed \cite{Jaksch2000}.  Each involves using the internal states of the atom as qubits.

An alternative is to use the center-of-mass degree of freedom, which can be manipulated by either using vibrationally excited states \cite{Eckert2002,Charron2002}, or lattices with less-than-unit filling \cite{Mompart2003}.  These proposals require careful control of the qubit states and operate in the limit of weak tunneling.  However, in a recent elegant experiment \cite{Anderlini2007}, both the spatial and internal degrees of freedom of two atoms were used to demonstrate the elements of a two-qubit exchange gate.  This experiment was in the regime of strong tunneling, bringing both atoms into the same well. Tunneling of single and pairs of atoms has been explored in an independent double-well experiment \cite{Foelling2007}.

In this paper, we show how a very simple quantum logic gate can be implemented using only the vibrational states \cite{Spielman2006,Mueller2007} of the recently realized double-well lattice \cite{Anderlini2006,Jenny2006,Anderlini2007}, and using simple control protocols well suited to experimental realization.  Central to this scheme is controlling the double wells to switch on the tunneling of an atom from one well to the other.  This tunneling, in turn, introduces collisions leading to an overall quantum phase.  Furthermore, by using the vibrational states this gate does not depend on the internal state, and thus requires no active stabilization of the magnetic field.  

\begin{figure}
\begin{center}
\includegraphics[width = 3.25 in]{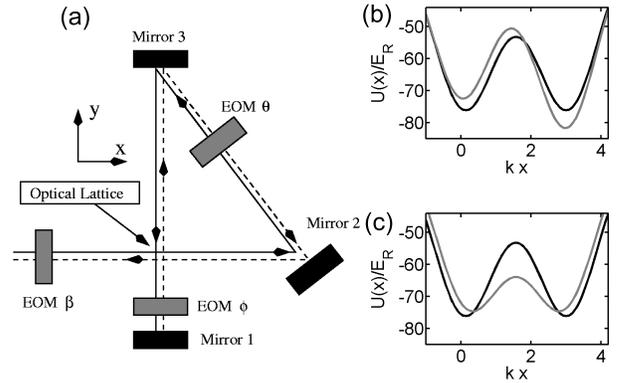}
\caption{(a) Two-dimensional optical lattice setup.  An additional optical lattice in the $z$-direction (out of the plane) provides three-dimensional confinement.  (b) The effective potential $U(x)$ (see text) for the double-well lattice with $\theta_z = \pi/2$ (black) and $\theta_z = \pi/2 + 0.5$ (gray) for $Z_f = 0.11$ and $V_0 = 40 E_R$.  The potential is plotted in units of the recoil energy $E_R$ and as a function of the dimensionless position $k x$.  (c) The effective potential $U(x)$ for the double-well lattice with $Z_f = 0.11$ (black) and $Z_f = 0.2$ (gray) for $\theta_z = \pi/2$ and $V_0 = 40 E_R$.}
\label{dwlattice}
\end{center}
\end{figure}

A two--dimensional array of double wells is created \cite{Jenny2006,Anderlini2006} by intersecting at right angles two pairs of counterpropagating laser beams, with an independent optical lattice along the orthogonal direction.  This is used to trap and manipulate cold $^{87}$Rb atoms.  The four beams are obtained by folding and retro--reflecting a single laser beam as shown in Fig.\ref{dwlattice}(a).  All four beams lie in a horizontal plane.  The electro--optic modulator labeled EOM $\beta$ is used to rotate the polarization of the incoming beam [solid line in Fig.\ \ref{dwlattice}(a)].  Consequently, the sum of the electric fields of the four beams in the lattice region has components in both the horizontal ($xy$, in-plane) and vertical ($z$,out-of-plane) directions. The EOM devices labeled $\theta$ and $\phi$ serve to introduce phase shifts ($\theta_{z}$ and $\phi_{z}$) between these two electric--field components by altering the optical path lengths.

The potential experienced by the atom is proportional to the total 
intensity of the light.  In this case the full optical potential is
\begin{eqnarray}
V\left(x,y,z\right) &=& -\frac{1}{4}V_{1}
\big[4 + 
2\cos\left(2ky+2\phi_{xy}\right)\nonumber\\
& & \quad + 2\cos\left(2kx-2\theta_{xy}-2\phi_{xy}\right)\big]\nonumber\\
& & -V_{2}\big[
\cos\left(kx-\theta_{z}-\phi_{z}\right) +
\cos\left(ky+\phi_{z}\right)
\big]^{2}\nonumber\\
& & + V_3 \sin^2 (k z),
\label{full_2d_potential}
\end{eqnarray}
where $V_{1}$ and $V_{2}$ are proportional to the fractions of the total light intensity contributed by the in--plane and out--of--plane polarized light, respectively, $V_3$ to the light intensity for the optical lattice along the $z$-axis, and $k = 2\pi/\lambda$, where $\lambda$ is the wavelength of the incident lasers.  It will be convenient in what follows to define $V_{0}$ and $Z_{f}$ such that $V_{1} = V_{0}\left(1-Z_{f}\right)$ and $V_{2} = V_{0}Z_{f}$.  The electro--optic elements enable control of the relative phases $\theta_{z}-\theta_{xy}$ and $\phi_{z}-\phi_{xy}$.  In this paper we shall only consider the case where $\theta_{xy}=\phi_{xy}=\phi_{z}=0$.  

When the depth $V_{0}$ of the lattice is sufficient, the system can be arranged so that each double--well lattice site contains exactly two
atoms.  In describing a single two--atom pair, 
we may consider a one--dimensional ``cut'' through the potential at 
$z=y=0$ and along the $x$ axis, $U(x)=V(x,0,0)$.  Using the three
parameters appearing in this potential, $\theta_{z}$, $Z_{f}$, and $V_{0}$, 
control can be exercised over the ``tilt'', the barrier height between the two wells, and the overall depth of the lattice.  The tilt, defined as the energy difference between the left and right well minima, is controlled mainly by changing $\theta_{z}$ as can be seen in Fig. \ref{dwlattice}(b).  Tuning the value of $Z_{f}$ changes the height of the central barrier, as seen in Fig. \ref{dwlattice}(c).  The phase gate operation proposed in this paper is performed by changing the value of $\theta_{z}$ only.

For a single atom in one site of the optical potential described by Eq. (\ref{full_2d_potential}), we encounter a number of energy scales.  By expanding the potential $V(x,y,z)$ in $y$ and $z$ we find $V(x,y,z) \approx U(x) + (1/2) m \omega_y^2 y^2 [1+Z_f \cos(kx - \theta_z)] + (1/2) m \omega_z^2 z^2$, with $\hbar \omega_y = \sqrt{4 E_R V_0}$ and $\hbar \omega_z = \sqrt{4 E_R V_3}$, where we have introduced the recoil energy $E_R = (\hbar k)^2/(2m)$ with $m$ the atomic mass.  For typical lattice parameters $Z_f = 0.11$, $V_3 = V_0 = 40 E_R$, we find that $\hbar \omega_y = \hbar \omega_z = 12.65 E_R$.  These are the approximate energies for excitation along the $y$ or $z$ directions, and are larger than an excitation in the double-well potential $U(x)$, which is approximately $10 E_R$.  Interestingly, the residual coupling between $x$ and $y$ does not lead to any first-order mixing of the states, but does shift the central barrier by $Z_f \sqrt{V_0/E_R}/4 \approx 0.2 E_R$, which is small compared to the typical barrier height of $20 E_R$ (see Fig.~\ref{dwlattice}), and thus can be neglected.  In the double-well potential, pairs of nearly degenerate levels (right and left) are actually split.  For the ground doublet this splitting is very small, but for the upper doublet, this tunnel splitting $2 J \approx 0.3 E_R$ ($2J/h$ is the tunneling frequency).  As our gate involves only manipulations of energy at the level of $J$, we will restrict our attention to one-dimensinal motion along the double-well axis ($x$). 

For two atoms, we must also consider the effective one--dimensional atom--atom interaction, which arises from the three-dimensional interaction $(4 \pi \hbar^2 a_s/m) \delta({\bf r}_1 - {\bf r}_2)$,
where $a_{s}$ is the s--wave scattering length of the freely scattering atoms.  By integrating the $\delta$-function over $y$ and $z$, the interaction strength $g_{1D}$ along $x$ can be calculated (more sophisticated methods are described in \cite{Naidon2007}).  Using the ground state harmonic oscillator wavefunctions for $y$ and $z$ we find $g_{1D} = 2 \hbar \sqrt{\omega_y \omega_z} a_s$.  Note that this value can be conveniently described by the dimensionless parameter $\bar{g}_{1D} = k g_{1D}/E_R = 8 \pi (a_s/\lambda) \sqrt{V_0 /E_R}$, which with typical experimental parameters ($a_{s} = 5.3$ nm, $\lambda = 810$ nm, and $V_{0}/E_{R} = 40$) leads to $\bar{g}_{1D} \approx 1$.  

\begin{figure*}[ht]
\begin{center}
\includegraphics[width = 6.5 in]{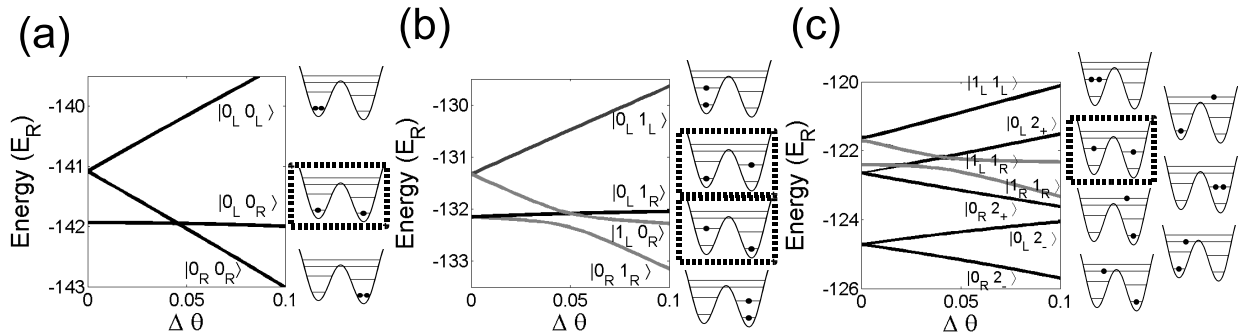}
\caption{Energy levels of the double-well potential, for $^{87}$Rb atoms in a double-well lattice as a function of $\Delta \theta = \theta_z - \pi/2$ with $V_0 = 40 E_R$, $Z_f = 0.11$, and $\bar{g}_{1D} = 1$.  Only those states with vibrational excitation along the direction of the double-well are calculated. While each apparent crossing of eigenvalues is avoided, the first-order tunneling splittings have been highlighted by the gray line segments in (b) and (c).  For each eigenvalue an approximate label ($|0_R 0_R\rangle$, etc.) indicates the eigenvector, valid for large $\Delta \theta$.  Next to each panel are schematic energy configurations for each eigenvalue that indicate the occupations of the single-particle energy levels.  The boxed configurations are the two-qubit states of the system.}
\label{dwlevels}
\end{center}
\end{figure*}

To design the two--qubit gate operated in this optical lattice, we now consider the energy--level spectrum of two interacting particles in the double--well potential, using the following Hamiltonian:
\begin{equation}
H = 
\frac{p_{1}^{2}}{2m} + U\left(x_{1}\right) +
\frac{p_{2}^{2}}{2m} + U\left(x_{2}\right) + 
g_{1D}\delta\left(x_{1}-x_{2}\right)
\label{2-atom_ham},
\end{equation}
where $p_{1}$ and $p_{2}$ are the momenta of atoms 1 and 2.  Each atom is assumed to move in a single period of the double--well potential such as that shown in Fig. \ref{dwlattice}.  

We propose to use as qubits the single--particle vibrational states supported 
by the double--well potential.  If the double--well potential is sufficiently tilted and has a high--enough central barrier, the lowest four single--particle motional--state energies will lie below the barrier and will divide into pairs of states with their wavefunctions well--localized on one or the other side of the barrier.  The key to our proposed two--qubit gate is to initially set the potential parameters such that these particular single--particle motional states exist and can be individually addressed.  Gate operation proceeds by varying the tilt (via the $\theta_{z}$ parameter) in such a way as to entangle the atoms, finally ending up with the same double--well potential.

With this goal in mind, we now calculate the eigenvalues of the two-particle Hamiltonian in Figure \ref{dwlevels} as a function of $\Delta \theta = \theta_z - \pi/2$ (proportional to the ``tilt'') with experimentally relevant parameters.  The energies of the first fourteen states are shown and labeled by their dominant composition (for large tilt) in terms of the products of the lowest six single-particle states of the double-well: $|0_R\rangle$, $|0_L\rangle$, $|1_R\rangle$, $|1_L\rangle$, $|2_-\rangle$, and $|2_+\rangle$.  The states subscripted by $L$ or $R$ are localized primarily in the left or right wells, while for these parameters $|2_-\rangle$ and $|2_+\rangle$ are not.  Also shown next to each of Fig. \ref{dwlevels}(a)-(c) are schematic configurations showing which of the single-particle levels are occupied.  The boxed configurations correspond to $|0_L 0_R\rangle$, $|1_L 0_R\rangle$, $|0_L 1_R\rangle$, and $|1_L 1_R\rangle$, the natural two-qubit states.  Note that in the above, Bose-symmetrization of the two-particle states has been suppressed to simplify the labels ({\it{e.g.}}, $|0_L 0_R\rangle \equiv 2^{-1/2} |0_L\rangle \otimes |0_R\rangle + 2^{-1/2} |0_R\rangle \otimes |0_L\rangle$).    

There are several things to observe in the eigenvalue spectrum.  First, for zero tilt ($\Delta \theta = 0$), there is complete symmetry between left and right.  Furthermore, those states with atoms localized in the same well are shifted up in energy.  For example, in Fig. \ref{dwlevels}(a) states $|0_L 0_L\rangle$ and $|0_R 0_R\rangle$ have greater energy than $|0_L 0_R\rangle$ at $\Delta \theta = 0$.  This difference in energy is the collisional interaction energy $U_{00}$ and is approximately one recoil energy ($E_R$) for these parameters.  Second, for sufficiently large tilt ($\Delta \theta \approx 0.1$), the qubit states are both spatially separated and separated in energy, so that single-qubit manipulation can be performed on each of the two atoms.  

Between these two extremes, at intermediate tilt, we observe several apparent intersections of the eigenvalue curves in Fig. \ref{dwlevels}.  Each of these is actually an avoided crossing, most due to second-order tunneling processes ($J^2/U$) as studied in a recent experiment \cite{Foelling2007}.  These splittings are very small and will not be used in the following.  Broader first-order tunnel splittings are found between states $|1_L 0_R\rangle$ and $|0_R 1_R\rangle$ in Fig. \ref{dwlevels}(b) and between states $|1_L 1_R\rangle$ and $|1_R 1_R\rangle$ in Fig. \ref{dwlevels}(c).  The corresponding energy levels have been highlighted in gray for emphasis.  Note that these two avoided crossings occur at slightly different tilts: the first occurs at $\Delta \theta \approx 0.05$, the second at $\Delta \theta \approx 0.04$.  They also have slightly different minimum splittings, which we now consider.

In a simple two-state model of each avoided crossing, the two splittings are approximately given by $\sqrt{(U_{01}-\Delta E)^2 + 4 J^2}$ and $\sqrt{(U_{11} - \Delta E)^2 + 8 J^2}$, where $2 J$ is the single-particle tunnel splitting (of the upper doublet; the second splitting is enhanced by a factor of $\sqrt{2}$ due to Bose symmetry), $\Delta E$ is the energy tilt, and $U_{01}$ and $U_{11}$ are the interaction energies for the two states $|0_R 1_R\rangle$ and $|1_R 1_R\rangle$, respectively.  Note that, to achieve maximal tunneling of the atoms, one must tilt the double wells to $\Delta E \sim U$, using the tilt to compensate for the interaction energy.  The two interaction energies are different; using a harmonic oscillator approximation, we find $U_{01} = U_{00}$, while $U_{11} = 3 U_{00}/4$.  However, by choosing the tilt such that $\Delta E \approx 7 U_{00}/8 - 8 J^2/U_{00}$, one can induce tunneling (from left to right) with the {\it same} oscillation frequency starting from either of the qubit states $|1_L 0_R\rangle$ or $|1_L 1_R\rangle$.  Both interactions and strong tunneling are required to achieve this type of sychronized oscillation. 

\begin{figure}
\begin{center}
\includegraphics[width = 3 in]{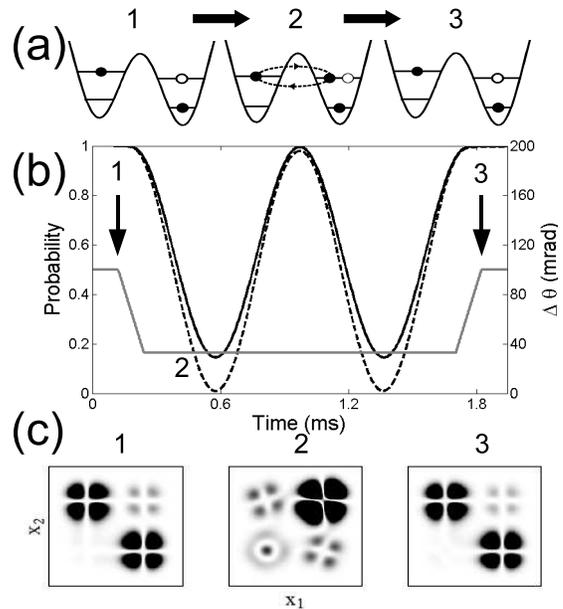}
\caption{(a) Schematic implementation of the tunneling phase gate consists of a three step sequence, from large ($1$) to small tilt ($2$) and back ($3$).  During the hold period (including time $2$), an atom in the first excited state of the left well will tunnel to and from the right well. (b) Time oscillations of the qubit populations given the initial condition $|\Psi(0)\rangle = |1_L 0_R\rangle$ (solid) and $|\Psi(0)\rangle = |1_L 1_R \rangle$ (dashed).  Also illustrated (gray solid) is the optical lattice parameter $\Delta \theta = \theta_z - \pi/2$ (right axis).  (c) Two-atom probability densities $|\Psi(x_1,x_2)|^2$, for various stages of the tunneling phase gate sequence, as indicated in (b), and the initial condition $|\Psi(0)\rangle = |1_L 1_R \rangle$.  }
\label{dwsequence}
\end{center}
\end{figure}

We now turn to the gate sequence, illustrated in Fig.\ref{dwsequence}.  Starting from large tilt (time $1$), in which the two--qubit states can be individually addressed, we shift $\Delta \theta$ to an intermediate tilt optimized such the tunneling frequencies are equal.  Holding the lattice at this tilt for some amount of time (including time $2$), collisions will occur when both atoms occupy the right well.  After some time an appreciable phase will develop, after which a tilt back will recover the original two-qubit states (time $3$), with an overall controlled phase.

The synchronized tunneling oscillations are demonstrated in Fig.\ref{dwsequence}(b), by solving the time-dependent Schr{\"o}dinger equation with the Hamiltonian (\ref{2-atom_ham}) using a split-operator spectral method and system parameters from Fig. \ref{dwlevels}.  Here the atom-atom interaction has been approximated by a superposition of two Gaussians, optimized to match the energy levels shown in Fig. \ref{dwlevels}.  Only the tilt $\Delta \theta = \theta_z -\pi/2$ is manipulated, using a ramp from $\Delta \theta_i = 100$ mrad to $\Delta \theta_h = 33$ mrad in time $t_r = 0.12$ ms, and a hold time of $t_h = 1.46$ ms, for an overall gate time of $t_g = t_h + 2 t_r = 1.7$ ms.  The two-particle probability density for various steps during this sequence is shown in Fig.\ref{dwsequence}(c).  To interpret these panels, observe that the initial state has little amplitude near the diagonal ($x_1 = x_2$), indicating that the atoms are well separated.  During the hold, when one atom tunnels from left to right, there is a large probability along the diagonal, indicating that both atoms are in the right well.  Finally, at the end, the wavefunction has returned to (approximately) the initial condition.  

After two oscillations of the probability density, the overall controlled phase that is accumulated (not shown) is $\phi = \phi_{11} + \phi_{00} - \phi_{01} - \phi_{10} \approx 0.9 \pi$ (this phase can be adjusted by changing the ramp parameters; see below).  To quantify the accuracy of this quantum logic operation, we construct the two-qubit gate's matrix representation $V$ by evolving each of the initial states corresponding to $|0_L 0_R\rangle$, $|1_L 0_R\rangle$, $|0_L 1_R\rangle$, and $|1_L 1_R\rangle$, and after removing single-qubit phases we calculate the average gate fidelity $\mathcal{F}$ \cite{Nielsen2002}:
\begin{equation}
\mathcal{F} = \frac{1}{5} + \frac{1}{80} \sum_{j,k=0}^{3} {\text{tr}}\left(W \sigma_{j}^{(1)} \sigma_{k}^{(2)} W^{\dagger} V \sigma_{j}^{(1)} \sigma_{k}^{(2)} V^{\dagger} \right)
\end{equation}
where $W = \text{diag}(1,1,1,e^{i\phi})$ and $\sigma_{j}^{(1)}$ are the Pauli matrices for the first qubit.  This is equivalent to an average of the gate fidelity over all initial two-qubit states.  

\begin{figure}
\begin{center}
\includegraphics[width = 3.25 in]{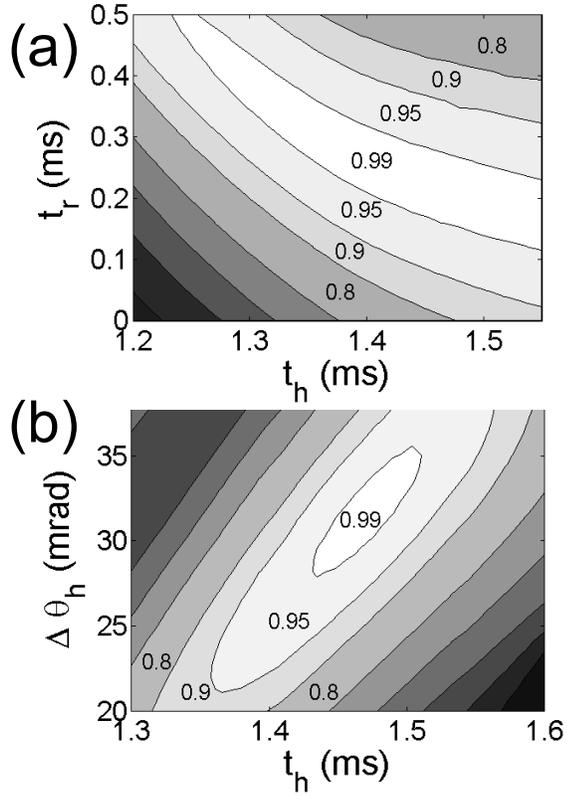}
\caption{(a) Average gate fidelity $\mathcal{F}$ as a function of the hold time $t_h$ and the ramp time $t_r$, with $\Delta \theta_i = 100$ mrad, $\Delta \theta_h = 34$ mrad, $V_0 = 40 E_R$, and $Z_f = 0.11$.  The labels between the contours indicate the minimum fidelity for the shaded regions.  The fidelity maximum is 0.994. (b) Average gate fidelity $\mathcal{F}$ as a function of the hold time $t_h$ and the final tilt $\Delta \theta_f$, with $t_r = 0.12$ ms, $\Delta \theta_i = 100$ mrad, $V_0 = 40 E_R$, and $Z_f = 0.11$.}
\label{dwfidelity}
\end{center}
\end{figure}

This fidelity is calculated for various values of the ramp and hold times $t_r$ and $t_h$ in Fig. \ref{dwfidelity}(a), and for various values of the final tilt $\Delta \theta_h$ and hold time in Fig. \ref{dwfidelity}(b).  Even for this simplest of control sequences, fidelities greater than 0.99 are possible.  Furthermore, by operating in the limit of strong tunneling, controlled by the tilt, we have achieved a gate operation that is nearly ten times faster than previous proposals \cite{Eckert2002,Charron2002}.  While promising, this gate is not quite optimal: a gate equivalent to a controlled--NOT requires $\phi = \pi$.  Improvement of both the controlled phase and overall fidelity should be possible by waiting longer, using tighter transverse confinement, a deeper lattice, or through optimal control techniques \cite{Calarco2004}.  Other issues that may limit the fidelity are coupling to vibrational excitations in the transverse directions, and will be studied elsewhere.  Nevertheless, Fig. \ref{dwfidelity} shows that small imperfections in the control parameters (due to timing errors, or inhomogeneities across the lattice) still allow for fidelities greater than 0.9, very promising for initial experimental demonstration.
 
In summary, we have analyzed the interaction of two neutral atoms in one cell of the double-well optical lattice.  By manipulating just one property of the potential, the ``tilt'', we have shown how to achieve a high-fidelity controlled phase gate.  By operating in the regime of strong tunneling, this scheme leads to a fast gate operation and is experimentally accessible without the need to control the internal states of the atom.  Consequently, this protocol can be performed without stabilizing the magnetic field environment, avoiding a significant source of decoherence.

\begin{acknowledgments}
\end{acknowledgments}

\end{document}